# Mathematical modeling of the synergetic effect between radiotherapy and immunotherapy

**Yixun Xing[2], Casey Moore[3], Debabrata Saha[3], Dan Nguyen[1,3], MaryLena Bleile[1,4], Xun Jia[1,3], Robert Timmerman[3], Hao Peng[1,3,*], Steve Jiang[1,3,*]**

[1] Medical Artificial Intelligence and Automation Laboratory, University of Texas Southwestern Medical Center, Dallas, TX 75390, USA
[2] Department of Advanced Data Analytics, University of North Texas, Denton, TX 76205, USA
[3] Department of Radiation Oncology, University of Texas Southwestern Medical Center, Dallas, TX 75390, USA
[4] Department of Statistical Science, Southern Methodist University, Dallas, TX 75275, USA

* **Correspondence:** Email: hao.peng@UTSouthwestern.edu, steve.jiang@utsouthwestern.edu.

**Abstract:** The synergy between radiotherapy and immunotherapy plays a pivotal role in enhancing tumor control and treatment outcomes. To explore the underlying mechanisms of synergy, we investigated a novel treatment approach known as personalized ultra-fractionated stereotactic adaptive radiation (PULSAR) therapy, which emphasizes the impact of radiation timing. Unlike conventional daily treatments, PULSAR delivers high-dose radiation in spaced intervals over weeks or months, enabling tumors to adapt and potentially enhancing synergy with immunotherapy. Drawing on insights from small-animal radiation studies, we developed a discrete-time model based on multiple difference equations to elucidate the temporal dynamics of tumor control driven by both radiation and the adaptive immune response. By accounting for the migration and infiltration of T cells within the tumor microenvironment, we established a quantitative link between radiation therapy and immunotherapy. Model parameters were estimated using a simulated annealing algorithm applied to training data, and our model achieved high accuracy for the test data with a root mean square error of 287 $mm^3$. Notably, our framework replicated the PULSAR effect observed in animal studies, revealing that longer intervals between radiation treatments in the context of immunotherapy yielded enhanced tumor control. Specifically, mice receiving immunotherapy alongside radiation pulses delivered at extended intervals, ten days, showed markedly improved tumor responses, whereas those treated with shorter intervals did not achieve comparable benefits. Moreover, our model offers an in-silico tool for designing future personalized ultra-fractionated stereotactic adaptive radiation trials. Overall, these findings underscore the critical importance of treatment timing in harnessing the synergy between radiotherapy and immunotherapy and highlight the potential of our model to optimize and individualize treatment protocols.

## 1. Introduction

The convergence of radiation therapy and immunotherapy has emerged as a pivotal approach in tumor management, resulting in improved patient outcomes and enhanced quality of life. This integration of advancements in physics and biology has led to the burgeoning field of radio-immunotherapy, with numerous ongoing trials [1–3]. However, these scientific advancements have also raised a host of new and significant questions. These include inquiries about the optimal radiation dosing scheme for stimulating an immune response, the timing of immunotherapy administration concerning radiation, and the integration of radiation therapy into the era of personalized precision medicine.

A novel paradigm, known as personalized ultra-fractionated stereotactic adaptive radiation (PULSAR) therapy, involves radiation delivery in ablative doses, with intervals spanning weeks or months, in contrast to the daily fractions commonly employed in clinical practice. The extended time between radiation doses enables adaptation to change within the tumor, allowing for a potential synergetic interaction with immunotherapies. Preliminary preclinical investigations into PULSAR in tandem with immunotherapy have underscored the impact of radiation scheduling on therapeutic efficacy. Notably, these studies revealed significant benefits when radiation was administered either as a single fraction or separated by ten days, whereas no similar benefit was observed for radiation pulses separated by just one day [4].

Although numerous prior studies have delved into the potential impact of radiation timing in standalone radiotherapy, such as stereotactic ablative radiotherapy (SABR), similar research in the realm of radioimmunotherapy remains relatively scarce. It is a challenging task that involves many complex biological processes. To name a few, irradiated T cells assume a critical role in controlling tumor growth post-radiation therapy. The tumor microenvironment may offer protection to irradiated T cells, which would otherwise undergo rapid demise elsewhere in the body [5]. Moreover, radiation may exert immune-inhibitory effects, a factor that must be carefully considered when pairing it with immunotherapy.

Our primary focus is to develop a mathematical model for the observed PULSAR effect in the experimental outcomes. The PULSAR effect encompasses not only improved tumor control with radiation pulses separated by one day, compared to a ten-day interval, in the absence of immunotherapy but also a reversed outcome in the presence of immunotherapy. Below, we provide a concise overview of several representative studies focused on modeling tumor growth and control. A top-down model was developed to illustrate tumor growth while coordinating radiotherapy with inhibitors targeting the PD1-PDL1 axis and the CTLA4 pathway [6]. This model was later simplified and adapted to explore the effects of single- and multiple-fraction schemes with 1-methyl tryptophan [7]. In another study, immune response during and after radiotherapy was modeled to analyze the growth of tumors in immune limited and immune escape modes [8]. Another macroscopic approach, considering tumor growth models (such as the Gompertz law) and the linear-quadratic (LQ), is presented in [9], where the interactions among the immune system, radiotherapy, and tumor progression are quantitatively analyzed. In contrast, the analytical model for chemoimmunotherapy in brain cancer offers a flexible, personalized approach by varying treatment intervals and dosages [10]. Similarly, a mathematical model was built to simulate and predict the response of non-small cell lung cancer patients to combined chemo- and radiotherapy using overall survival data from clinical trials [11]. Moreover, the researchers in [12] use clinical data analysis and simulation to compare the radiotherapy before and after the surgery in early-stage cancers, suggesting that preoperative radiation may elicit a more robust antitumor immune response. A three-dimensional agent-based model [13] simulates diverse tumor-immune ecosystems and introduces the individual radiation immune score as a predictor of radiocurability across multiple tumor types. Additionally, the model introduced in [14] is calibrated with experimental data to predict and dissect immune-mediated responses at multiple tumor sites following focal irradiation combined with systemic immunotherapy, identifying an optimal radiation dose range to maximize anti-tumor immunity. Kosinsky et al. [15] developed a quantitative pharmacologic model that illustrates the cancer immunity cycle, incorporating radiation treatment and therapeutic blockade of PD1 or PDL18. Despite its comprehensiveness, this model may face challenges with overfitting due to the numerous parameters involved. Such an approach was also applied to capture tumor dynamics and predict median clinical responses to monotherapy, combination, and sequential therapy involving the blockade of inhibitory effects by CTLA4, PD1, and PDL1 [16]. A bottom-up computational model was utilized to examine the tumor response specifically to anti-PD1 antibodies, employing just four parameters [17]. Additionally, Sung et al. proposed a quantitative model outlining the immunosuppressive and immune-stimulating effects induced by radiation therapy [18]. However, this model focused on simulating fractionation strategies without considering any immunotherapy modalities. Further extending these methodologies, the mechanistic framework in [19] utilizes ordinary differential equations to model the combined effects of immune checkpoint inhibitors and radiotherapy, particularly in hepatocellular carcinoma.

These results are promising, and researchers have explored PULSAR effect using deep learning models [20] or mathematical equations [21] based on limited experimental data. Nonetheless, determining the optimal fractionation schedule in combination with immunotherapy remains a relatively unexplored area, particularly within the PULSAR framework. Consequently, it is imperative to elucidate the intricate mechanism underlying PULSAR and optimize its synergy with immunotherapy accordingly. In this study, we developed a discrete-time mathematical framework and validated its efficacy using experimental data from mouse studies, considering the migration and infiltration of T cells as a function of both dose and time.

In Section 2, we introduce our mathematical framework along with the biological hypotheses upon which it is based. In Section 3, we detail the materials and methods used to fit and validate the model. In Section 4, we outline the experimental findings on the PULSAR effect and illustrate how our model reflects these outcomes, followed by additional in-silico predictions for various scenarios. A discussion of the model's implications and limitations is given in Section 5, with concluding remarks in Section 6.

## 2. Mathematical model

### *2.1. Biological hypotheses underlying the PULSAR effect*

In designing our mathematical framework, we focus on a set of biological assumptions to develop a concise system of equations. Incorporating the entirety of current immunology and radiotherapy knowledge would have resulted in an overly complex and unwieldy model with diminished limited practical applicability. Consequently, our model comprises just six equations, yet they effectively capture the key experimental outcomes of interest, especially the PULSAR effect.

- We assume that, in the absence of treatment or an immune response, the tumor grows at a constant exponential rate. Although more complex models exist, this simple exponential assumption is sufficient to capture the behavior of small murine tumors [6,7].
- We adopt the standard linear-quadratic model to characterize radiation-induced tumor cell damage. To account for the big tumor control penalty observed with longer intervals between radiation treatments in the absence of immunotherapy [4], we assume that the post-irradiation radiosensitivity decays on a daily basis.
- It is assumed that effective radio-immunotherapy relies on distinguishing two types of T cells: Resident intratumoral T cells, which are more radio-resistant, and newly infiltrating T cells that are recruited following radiation and can gradually convert into intratumoral T cells [4]. Studies show that although radiation damages T cells, many survive clinically relevant doses and display enhanced motility and immune activity [5]. Longer intervals between radiation pulses, in the presence of immunotherapy, improve T-cell recruitment and activation, thereby enhancing tumor control. Accordingly, our model incorporates T-cell migration and infiltration as functions of both dose and timing.

### *2.2. Model equations and interpretations*

Our discrete-time model captures the dynamics of three populations: Tumor cells and two types of T cells, as illustrated in Figure 1. Through a chain reaction of biological pathways, such as STING, radiation therapy (RT) recruits new T cells to the tumor site and stimulates their filtration, while killing both a subset of T cells and tumor cells. Additionally, tumor growth is modulated by effector T cells, whose activity can be enhanced by the checkpoint inhibitor antibody (anti-PD-L1) or suppressed by a monoclonal antibody (anti-CD8). Equations (1)–(6) outline the process governing these interactions and describe how radiation and immunotherapy jointly influence tumor progression in discrete time steps. Below, we provide a description of Eqs (1)–(6). For an in-depth explanation and additional details on Eqs (1)–(3), please refer to Section 2 of the supplemental materials.

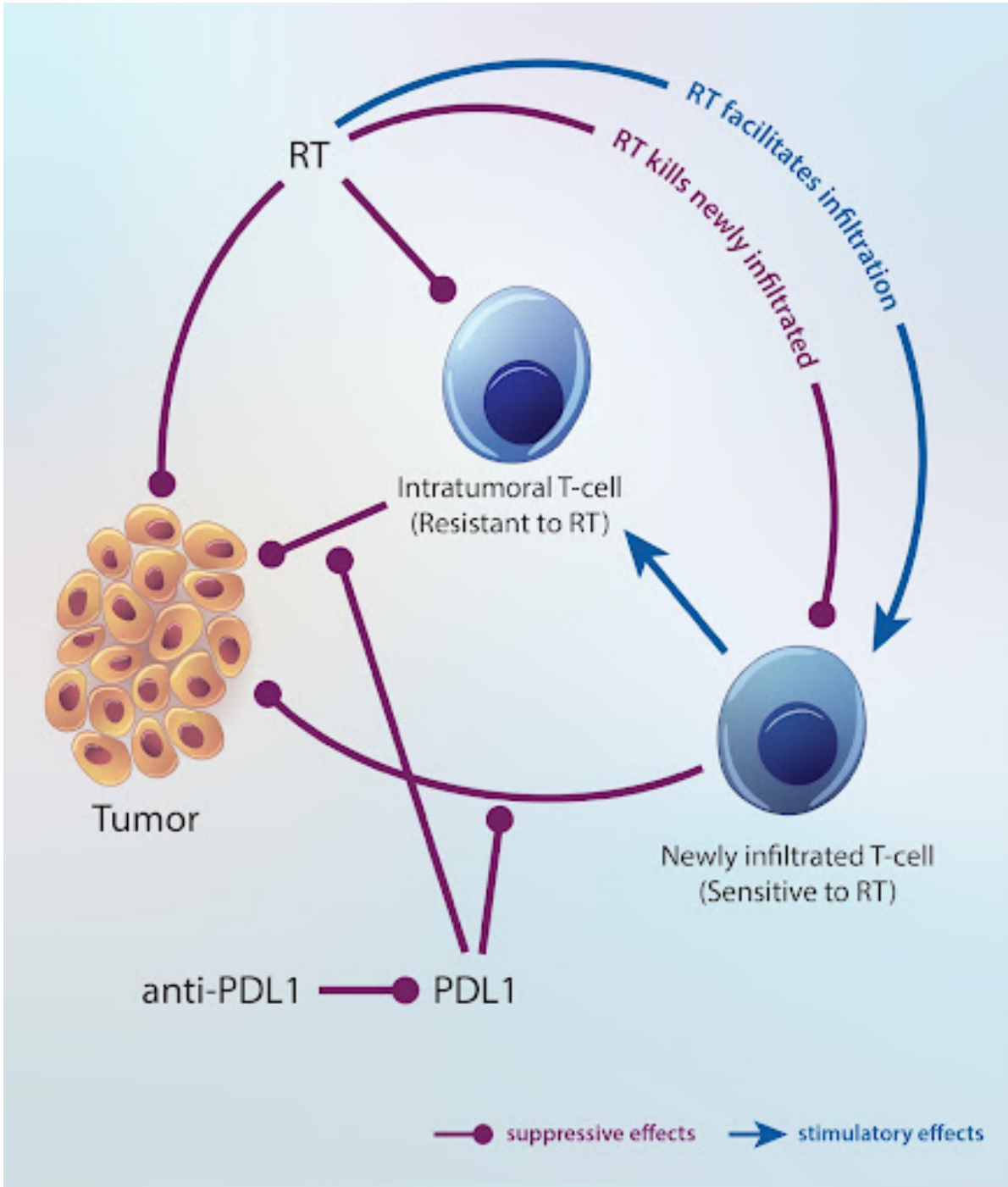

**Figure 1.** The interactions between the populations with immunosuppressive (red dot-headed lines) and stimulatory (blue arrows) effects.

**Tumor dynamics**

$$T_{n+1} = S_n T_n e^{\mu - Z_n} \tag{1}$$

The tumor growth dynamics follow Eq (1), where $T_n$ is the tumor volume on day n and grows exponentially to $T_{n+1}$ at a rate $\mu$ in the absence of treatment. $S_n$ represents the probability that tumor cells survive radiation on day $n$. $Z_n$ denotes the anti-tumor effect due to immune response (details in Eqs (4)–(6)). Radiation lowers tumor volume through $S_n$ while the immune system further reduces it via $e^{-Z_n}$.

**Survival probability under radiation**

$$S_n = e^{-(\alpha d_n + \beta d_n^2)} e^{-\gamma R_n (\alpha d_n + \beta d_n^2)} \tag{2}$$

$d_n$ represents the radiation dose on day $n$ while $\alpha$ and $\beta$ are the linear-quadratic parameters. $\gamma$ represents the scaling coefficient for accumulated radiosensitivity. $R_n$ denotes the cumulative radiosensitivity post-irradiation (details in Eq (3)). This equation extends the standard LQ model by introducing a 'memory-modulation' term, which enhances radio sensitivity based on prior radiation (more explanation in the Discussion section).

**Accumulated radiosensitivity**

$$R_{n+1} = min\,(\tau R_n + (1 - S_n),1) \tag{3}$$

Here, $\tau$ is the daily decay rate of $R_n$, the accumulated radiosensitivity through day *n-1*. Moreover, $(1 - S_n)$ represents the additional contribution from the radiation dose administered on day $n$ to the accumulated effect. Consequently, starting from $R_0 = 0$ and capped at a maximum of 1, $R_{n+1}$ accumulates radiation effect up to day $n$ and will carry it forward to day *n+1*. The terms $(1 - S_n)$ and $\tau R_n$ indicate that a higher dose or shorter interval between doses raises $R_{n+1}$, thereby enhancing radiosensitization for subsequent treatments.

**Anti-tumor T cell effect**

$$Z_n = \begin{cases} \omega_1 W_{tum,n} + \omega_2 W_{new,n} & if\ within\ 7\ days\ post\ anti\text{-}PD\text{-}L1 \\ 0 & otherwise \end{cases} \tag{4}$$

Here, $W_{tum,n}$ and $W_{new,n}$ represent the populations of existing intratumoral and newly infiltrated T cells, respectively, on day $n$. $\omega_1$ and $\omega_2$ denote the potency factors for each T cell population. This piecewise definition activates T cell-mediated anti-tumor effects for seven days after anti-PD-L1 administration, modeling a transient phase of heightened immune response.

**T cell dynamics**

$$W_{tum,n+1} = \lambda W_{new,n} + W_{tum,n} e^{-\phi_1 d_n} \tag{5}$$

$$W_{new,n+1} = \begin{cases} (1 - \lambda) W_{new,n} e^{-\phi_2 d_n} + \rho & if\ radiation\ is\ applied \\ (1 - \lambda) W_{new,n} e^{-\phi_2 d_n} & otherwise \end{cases} \tag{6}$$

$\phi_1$ and $\phi_2$ are the radiation sensitivity parameters for intratumoral and newly filtrating T cells, respectively. The conversion rate from $W_{new}$ to $W_{tum}$ is $\lambda$. $\rho$ represents the rate at which radiation recruits new T cells (assumed constant here for simplicity). Equations (5) and (6) define how each T cell population evolves under combined radiation and immunotherapy. It is expected that $\phi_1 < \phi_2$ because the intratumoral T cells are more radiation-resistant [5]. Also, we focus on immediate effects that manifest rapidly after radiation therapy (e.g., within hours than days). The generation of T cells resulting from radiation therapy is a complex process that can vary widely depending on several factors. Some tumors may be more immunogenic, leading to a stronger T cell response after radiation. Alternatively, a higher dose may stimulate the release of more tumor-associated antigens and lead to the recruitment of more T cells. For simplicity in this study, the infiltrating T cells are replenished following radiation through the constant $\rho$.

## 3. Materials and methods

### *3.1. Small animal experiments*

All animal experimental protocols, including those for mouse use and euthanasia, were reviewed and approved by the Institutional Animal Care and Use Committee (IACUC) of the University of Texas Southwestern (UTSW) Medical Center under animal protocol number 2018-102620. Female

C57BL/6J mice aged six to eight weeks were purchased from Charles River or Jackson Laboratories. Lewis lung carcinoma (LLC) was derived from lung cancer of the C57BL/6 line, and tumor cells were injected subcutaneously on the right leg of mice. Mice in each group were administered (i.p) α-PD-L1 (200 ug) or anti-CD8 agent (200 ug) with different schedules (see Figure A1). Tumor-bearing mice were anesthetized with isoflurane and irradiated with 10 to 40 Gy according to different schedules. Local irradiations were conducted on a dedicated X-ray irradiator (X-RAD 32, Precision X-ray, Inc.). Various collimator sizes were developed to form the field of view, depending on tumor size. The mouse was positioned so that the source-to-tumor surface distance was 20 cm, with the tumor placed at the center of the X-ray beam. The energy of the x-ray was set to 250 kVp, and the current was set to 15 mA for irradiation. The dose rate under this condition was 19.468 Gy/min, which was calibrated using a PTW 31010 ionization chamber and a PTW UnidosEelectrometer (PTW North America Corporation, New York, NY) following the AAPM TG-61 protocol.

The mice were randomized to treatment groups when tumors reached 150 to 200 mm$^3$. Figure A1 in the Appendix details the schedule for radiotherapy and immunotherapy, along with the names of corresponding treatment groups. The group name 'XGydXdX' encodes the treatment protocol: The first 'X' indicates the radiation dose; 'Gy' denotes radiation treatment; the first 'dX' marks the day of the initial radiation; and a second 'dX' (if present) marks the day of a subsequent radiation. If α-PD-L1 is administered, '+PDL' is appended. Generally, an initial dose of α-PD-L1 or isotype control was applied two or three days before the first radiation, on the day of the initial radiation, and then subsequently every other day for a period. The tumor volumes were measured by length (x), width (y), and height (z) and calculated as tumor volume = xyz/2. When the tumor volume exceeded 1500 mm$^3$, or if the mouse exhibited significant ulceration in the tumor, it reached the survival endpoint and was euthanized. The censored data were then processed using an imputation method [22,23], after which the mean of each treatment group was computed.

### 3.2. *Model fitting*

The above experimental data were divided into training and testing datasets, each comprising twelve distinct treatment groups, with schedules detailed in Figure A1 in the Appendix. The training data were used to estimate the model parameters, while the testing data were held aside for model validation. The model fitting procedure and simulations were implemented in R. Simulated annealing [24] was applied to the training data to optimize model parameters by minimizing the mean square error (MSE) loss function. Here, MSE is defined as the average of the squared differences between the model-predicted tumor volumes and the observed tumor sizes, computed on the days when experimental measurements were taken. In each iteration of the simulated annealing optimization process, Eqs (1)–(6) were sequentially evaluated to generate daily tumor volume estimates, with the free parameters updated accordingly.

At the beginning of the fitting process, parameters were initialized using multiple sets of starting values, each constrained within reasonable ranges. Table 1 summarizes the best-performing initial values, as measured by the minimum MSE on the training data, along with the corresponding search space ranges. We also carefully determined the initial values for all variables requiring initialization. Since $R_n$ represents the accumulated radiosensitivity, we initialized it with $R_0 = 0$. Although tumor volume could be measured after tumor inoculation, the actual initial tumor volume was 0 mm$^3$ for each mouse and so unmeasurable. However, our discrete-time model required a nonzero starting tumor volume, so we set $T_0 = 1\,mm^3$ for all groups. Furthermore, recognizing that actual initial volumes vary among mice, we adjusted $T_0$ for each group within a range of 1–10 mm³ to identify the value that minimized the MSE, thereby accounting for both inter-group and inter-animal variations. Additionally, since the initial number of resident intratumoral and newly infiltrating T cells can differ from zero, we also estimated the initial values $W_{tum,0}$ and $W_{new,0}$ during the training phase. We initialized $W_{tum,0}$ and $W_{new,0}$ to 2 and 1, respectively, and then searched within a range from 1 to 100 to determine the optimal values that minimized MSE at the end of training.

The iterative fitting process continued until convergence was achieved. We evaluated three distinct fitting procedures, as described below:

1) Modeling with predetermined and fixed values of $\alpha$ and $\beta$ at 0.23595 and 0.036284, respectively. This was followed by estimating all other parameters.

2) Estimating all parameters by fitting groups sequentially in three stages. First, we fitted the tumor growth rate $\mu$ within the 0Gy group without anti-PD-L1 and $T_0$. Second, with $\mu$ fixed from the previous stage, $\alpha$, $\beta$, $\tau$, and $\gamma$, along with group-wise $T_0$, were optimized from training data of regimens involving only RTs, i.e., groups 10Gyd0d1, 10Gyd0d10, 10Gyd0d1d10d11, 10Gyd0d20, and 10Gyd0d1d20d21, respectively. In the final optimization step, we used the above estimates of $\alpha$, $\beta$, $\tau$, and $\gamma$ as initial values. $W_{tum,0}$, $W_{new,0}$, $T_0$, and all other parameters, except the fixed $\mu$, were assessed using all available data except for the 0Gy group.

3) Simultaneously estimating all parameters by fitting all treatment groups.

It is important to highlight that we discovered that the simultaneous estimation of all parameters using the complete dataset yielded superior results when compared to two alternative methods. Therefore, only the results of the simultaneous estimation are presented in this manuscript.

**Table 1.** Parameter space and initial values.

| Parameter | Lower | Upper | Initial value for optimization |
|---|---|---|---|
| $\alpha$ | 0 | 1 | 0.01 |
| $\beta$ | 0 | 1 | 0.0046 |
| $\omega_1$ | 0 | 1 | 0.0005 |
| $\omega_2$ | 0 | 1 | 0.0072 |
| $\phi_2$ | 0 | 1 | 0.946 |
| $\lambda$ | 0 | 1 | 0.205 |
| $\rho$ | 0 | 100 | 4.3 |
| $\phi_1$ | 0 | 1 | 0.349 |
| $\mu$ | 0 | 1 | 0.1813 |
| $\gamma$ | 0 | 10 | 0.15 |
| $\tau$ | 0 | 0.9 | 0.9 |

### *3.3. In silico outcome simulation*

Our model has the potential to inform future design of PULSAR trials. As an example, we simulated tumor growth under twelve different treatment regimens after optimizing parameters and initial variable values following the procedure described in Section 3.2. These regimens consist of two radiation pulses delivered at intervals of 4, 8, or 12 days, with doses of either 15 Gy or 20 Gy per pulse, with or without anti-PD-L1. The initial tumor size was set uniformly across all groups, using the average of the adjusted initial tumor volumes, 2.9 mm$^3$, from the training data, while $W_{tum,0}$, $W_{new,0}$ and all other parameters were maintained at the values, as described in Section 4.2 and Table 2.

## 4. Results

### *4.1. Experimental results and PULSAR effect*

To illustrate the PULSAR effect, in this section, we summarize key findings from the experimental dataset, including training and testing data. In Figures 2 and 3, the blue curves—solid for groups receiving immunotherapy and dashed for those without—highlight prominent trends in tumor growth across different radiation timing intervals. By focusing exclusively on these experimental blue lines, we draw attention to the distinct outcomes observed with and without immunotherapy, thereby emphasizing the impact of treatment timing on tumor control. Detailed quantitative results of the model's performance are provided in Sections 4.2–4.4.

20Gyd0, 10Gyd0d1, 10Gyd0d10, and 10Gyd0d4: All these treatment regimens received an identical radiation dose, yet different interactions with anti-PD-L1 were observed for the radiation fractions administered at different time intervals. An intriguing observation was the divergence between the solid and dashed blue lines in 20Gyd0 (receiving the highest anti-PD-L1), 10Gyd0d10, but not in the case of 10Gyd0d4 or 10Gyd0d1. The observed PULSAR effect was evident when comparing groups 10Gyd0d10, 10Gyd0d4, and 10Gyd0d1. In these groups, mice that received immunotherapy in conjunction with radiation pulses delivered at extended intervals (e.g., 10 days) exhibited significantly improved tumor responses, whereas those treated with shorter intervals (1 or 4 days) did not show comparable benefits. As hypothesized in Section 2.1, immunologically significant processes that normally occur between days 1 and 10 may be disrupted by an additional radiation dose, thereby reducing the effectiveness of anti-PD-L1 therapy. Additionally, when comparing the 10Gyd0d1 and 10Gyd0d10 groups in the absence of immunotherapy (illustrated by the solid blue lines in the top right and middle left panels of Figure 2), a pronounced tumor control penalty was observed with the 10-day interval (10Gyd0d10) relative to the 1-day interval (10Gyd0d1).

20Gyd0d10 and 15Gyd0d10: An additional advantage of immunotherapy as apparent when a higher radiotherapy dose was administered at a 10-day interval. The two curves exhibited a striking similarity, while the 15Gyd0d10 group showed relatively inferior tumor control. This extended timeframe enabled enhanced immune cascades and the arrival of different types of cells to signal, enhancing the overall anti-tumor immune response.

10Gyd0d10 and 10Gyd0d1d10d11: The increased radiotherapy dose should typically result in better tumor control. In the 10Gyd0d1d10d11 group, the tumor had a 10-day interval between radiation treatments but received a higher cumulative radiation dose, always followed by a second dose the day after the first. However, the difference between the solid blue and dashed blue line was less in the 10Gyd0d1d10d1 group relative to the 10Gyd0d10 group. In our view, this was likely due to the infiltration of some immune cells into the tumor within 24 hours after radiation, which were subsequently eliminated by the second dose administered on day 1.

(10Gyd0d10 and 10Gyd0d20): The PULSAR effect was evident in the former scenario but not in the latter. In the 10Gyd0d20 group, there was a less pronounced benefit from additional PD-L1 when the administration was delayed for 20 days. This was likely because the tumors were growing rapidly, and it appeared that an additional radiation dose was necessary to bring them under control in time.

Overall, there appeared to be a delicate balance between reaching an equilibrium that fosters synergy and enabling radiotherapy to play a more proactive role in managing tumor growth.

10Gyd0d1d10d11, 10Gyd0d1d20d21, and 40Gyd0: All three regimens delivered the same cumulative radiation dose. Setting aside toxicity and side effects, the most effective tumor control was observed in the 40Gyd0 group, where radiotherapy was administered all at once. The PULSAR effect was observed except in the 10Gyd0d1d20d21 group. This implied either that the immune response to a single higher dose as superior to lower doses or that the subsequent 3x10Gy treatments worsened the outcome.

Below are two additional observations to note. First, in all the figures, the clear separation between the solid blue and dashed blue lines did not occur until around day 10. This delay in response was expected in LLC, a tumor that was inherently resistant to immunotherapy, which highlighted the role of radiotherapy in altering the tumor's immune microenvironment and triggering cascades of immune responses. The 10-day lag also suggested the involvement of lymph nodes, as the priming of a new immune response through a lymph node typically took around 7–10 days. Furthermore, the difference between 15Gyd0d10 and 15Gyd0d10-CD8 confirmed CD8+ T cell involvement in the anti-tumor immune response, thus validating the efficacy of our model based on T cell dynamics.

## *4.2. Fitted model parameters*

As described in Section 3.2, we applied a simulated annealing algorithm to fit our mathematical model (Eqs (1)–(6)) to the training data from 12 treatment groups (depicted by the blue solid and dashed lines in Figure 2). All parameters were estimated simultaneously to minimize MSE during this process, and Table 2 summarizes the resulting estimates. Moreover, the initial values for both types of T cells, $W_{tum,0}$ and $W_{new,0}$, were close to zero.

The estimated daily tumor growth rate ($\mu$) was 0.216, which, as hypothesized in Section 2.1, adequately described the natural exponential growth of untreated tumors (0 Gy), as illustrated by the blue solid line in the first panel of Figure 2. The estimated nonzero value of $\rho$, 1.7, indicated that radiation effectively recruited new T cells. Morevoer, parameter $\phi_1$ was significantly lower than $\phi_2$, highlighting the different radiosensitivities of tumor-resident T cells versus newly infiltrated T cells, as proposed in Section 2.1. This estimation supported the model's ability to predict tumor volumes that were consistent with the observed PULSAR effect discussed in Section 4.1. Additionally, parameters $\omega_1$ and $\omega_2$ represented the contributions of the two T cell subpopulations, with $\omega_1$ being higher than $\omega_2$ —suggesting that intratumoral T cells are more responsive to immune effectors than newly infiltrating T cells. Moreover, the conversion parameter $\lambda$ was estimated at 0.304, indicating a moderate rate at which newly infiltrated effector cells became intratumoral [4,5]. Finally, the decay rate $\tau$ was estimated at 0.886, meaning that the accumulated post-irradiation radiosensitivity decreased by approximately 10% per day in the absence of additional radiation. This finding was consistent with experimental observations of $\tau$ reduced tumor control when the interval between radiation pulses was extended, as described in Section 4.1. A more in-depth explanation of these parameters can be found in the discussion section.

**Table 2.** Summary of fitted model parameters.

| Parameter | Estimate |
|---|---|
| $\boldsymbol{\alpha}$ | 0.0240 |
| $\boldsymbol{\beta}$ | 0.00148 |
| $\boldsymbol{\omega_1}$ | 0.0235 |
| $\boldsymbol{\omega_2}$ | 0.0140 |
| $\boldsymbol{\phi_2}$ | 0.964 |
| $\boldsymbol{\lambda}$ | 0.304 |
| $\boldsymbol{\rho}$ | 1.707 |
| $\boldsymbol{\phi_1}$ | 0.05205 |
| $\boldsymbol{\mu}$ | 0.216 |
| $\boldsymbol{\gamma}$ | 0.883 |
| $\boldsymbol{\tau}$ | 0.886 |

### *4.3. Modeling results on training and testing data*

#### *4.3.1. Training data results*

The fitted parameters described in Section 4.2 and Table 2 yielded optimal performance, achieving a root mean squared error (RMSE) of 320 $mm^3$ on the training data. Figure 2 displays the 12 treatment groups in the training dataset, comparing the experimental tumor growth data with the corresponding model predictions. Figure 2 consists of six panels, each representing a paired treatment group defined by a distinct radiation treatment with or without immunotherapy. In each panel, the dashed lines indicate the group receiving immunotherapy, while solid lines represent the corresponding group without immunotherapy. Two colors are used to show the source of tumor volumes: Blue lines represent the observed experimental data, and red lines denote the associated model predictions. Consequently, each panel features four curves: Solid blue (observed data without immunotherapy), dashed blue (observed data with immunotherapy), solid red (model predictions without immunotherapy), and dashed red (model predictions with immunotherapy). This arrangement facilitates a direct comparison of the model's performance against the experimental results across radiation timing scenarios. Our primary focus at this stage was on accurately replicating the observed PULSAR effect than merely optimizing quantitative metrics. Figure 2 demonstrates a strong alignment between the experimental training data and the predictions generated by our fitted model parameters, especially in capturing the key characteristics of the PULSAR effect.

0Gy: When comparing the solid blue and red lines, the experimental data and model predictions showed high consistency. For instance, on day 23, the experimental 0 Gy tumor volumes were 1605 $mm^3$ (without anti-PD-L1, solid blue) and 1706 $mm^3$ (with anti-PD-L1, dashed blue), while the model predicted 1430 $mm^3$ in both cases (solid and dashed red lines). This alignment indicated that the estimated daily tumor growth rate, $\mu$ (as described in Section 4.2), effectively captured the natural exponential growth of tumors in the absence of treatment or with only anti-PD-L1.

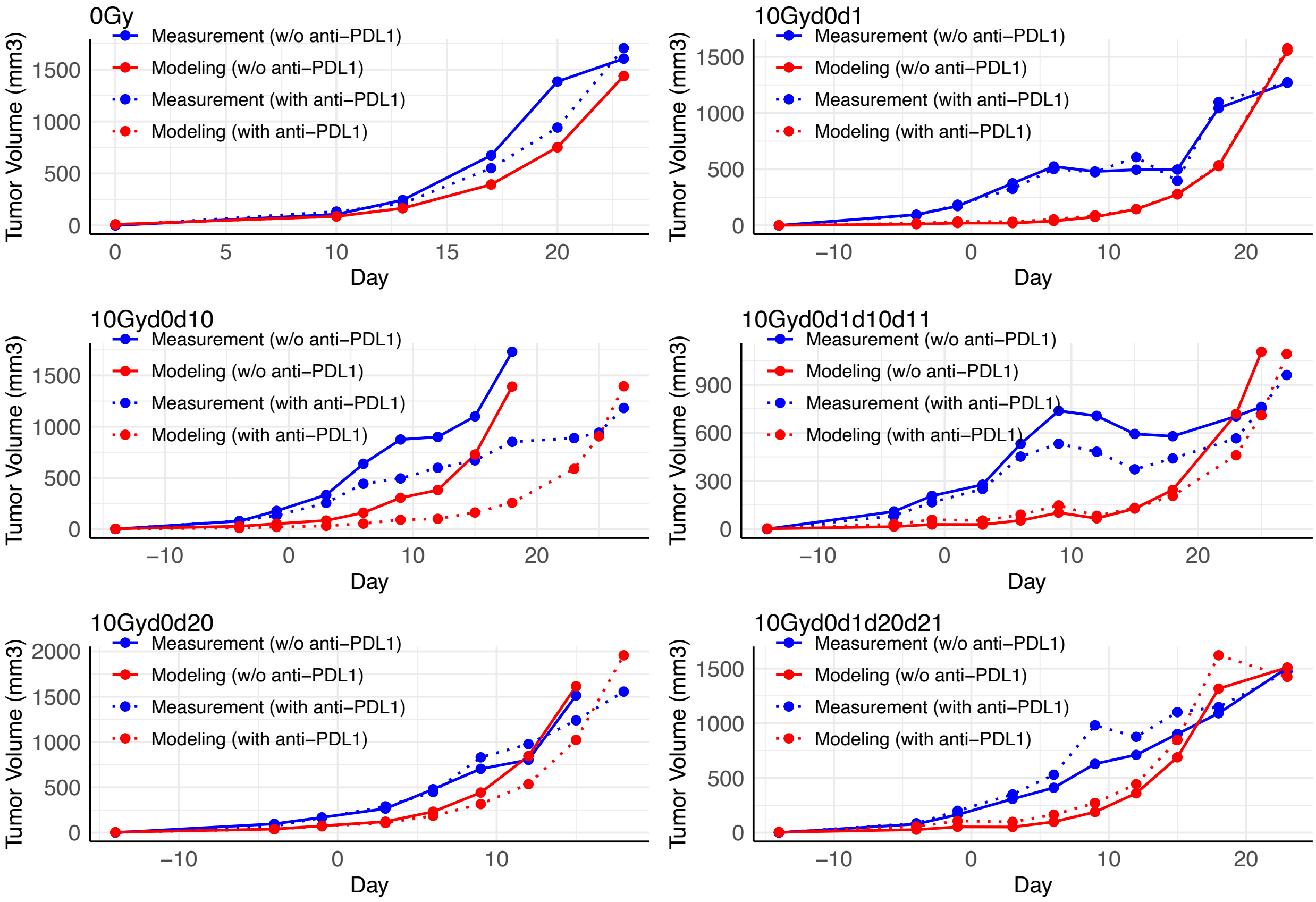

**Figure 2.** Experimental data (blue) used to estimate the parameters vs model-based tumor growth (red) of groups of different time spacing between the two fractions. In all the above plots, the solid line represents radiotherapy only, while the dashed line represents radiotherapy and immunotherapy combined. The 'Day' on the x-axis represents the day after the first RT if at least one radiation pulse is administered.

10Gyd0d1, 10Gyd0d10, and 10Gyd0d20: When comparing the solid blue and red lines, there wwas a consistency between experimental and model-based tumor growth. The measured tumor volumes on day 15 were 497 mm$^3$ (the 10Gyd0d1 group), 1100 mm$^3$ (the 10Gyd0d10 group), and 1514 mm$^3$ (the 10Gyd0d20 group), while their respective simulated counterparts were 276 mm$^3$, 728 mm$^3$, and 1615 mm$^3$. Moreover, when comparing the values of 10Gyd0d1 and 10Gyd0d10 groups, it was shown that longer radiation intervals in the absence of anti-PD-L1 led to poorer tumor control. We captured this difference via the accumulated radiosensitivity $R_n$ that decayed by $\tau$ (about 10% per day), so a shorter interval preserved a stronger effect for tumor control. More importantly, the modeling results underscored the most prominent efficacy of anti-PD-L1 when there was a ten-day gap between the two radiation fractions. The model's ability to capture treatment dynamics was largely attributable to the estimated values of $\phi_1$ and $\phi_2$, which accurately reflected the different radiosensitivities of the two T cell populations, as discussed in Section 4.2. In cases where the fraction interval of the split course was either just one day or as long as 20 days, the benefit from anti-PD-L1 was minimal. For the 10Gyd0d10 with anti-PD-L1 group, the simulated tumor volume on day 18 was 257 mm$^3$, which was significantly lower than the 1392 mm$^3$ for the group without anti-PD-L1. Conversely, the modeling outcomes of 280 mm$^3$ (the 10Gyd0d1 with anti-PD-L1 group) and 1023 mm$^3$ (the 10Gyd0d20 with anti-

PD-L1 group) on day 15 were close to 276 mm$^3$ (the 10Gyd0d1 group) and 1615 mm$^3$ (the 10Gyd0d20 group), respectively.

10Gyd0d1d10d11,10Gyd0d1d20d21: The estimated tumor volumes generally aligned with the actual ones. On day 23, the observed and estimated tumor volumes were 703 mm$^3$ and 718 mm$^3$ for the 0Gyd0d1d10d11 group and 1506 mm$^3$ and 1508 mm$^3$ for the 0Gyd0d1d20d21 group. Comparing these two groups without anti-PD-L1 further highlights the role of accumulated radiosensitivity—captured by $R_n$ and its decay rate $\tau$. Specifically, when the interval between the second and third pulses in the absence of immunotherapy was extended, the final tumor volume increased, suggesting that a longer gap leads to reduced cumulative radiosensitivity and, consequently, poorer tumor control. Furthermore, the model predicted a tumor volume of 460 mm$^3$ in the 10Gyd0d1d10d11 with the anti-PD-L1 group, showcasing the PULSAR effect.

#### *4.3.2. Testing data results*

When evaluated on a testing dataset, the model exhibited strong accuracy, achieving an RMSE of 287 mm$^3$. Figure 3, arranged similarly to Figure 2, displays the 12 treatment groups in the testing dataset. The patterns observed in Section 4.1 and the training data fitting results were consistently reproduced across testing groups with varying time intervals.

10Gyd0d4, 15Gyd0d10, 15Gyd0d10-CD8, and 20Gyd0: The modeling curves approximated the measurement curves well in these groups. The tumor volumes were 1011 mm$^3$ (the 10Gyd0d4 group), 1294 mm$^3$ (the 15Gyd0d10 group), 969 mm$^3$ (the 15Gyd0d10-CD8 group), and 1009 mm$^3$ (the 20Gyd0 group) on day 22, while the corresponding modeling results were 1015 mm$^3$, 1104 mm$^3$, 930 mm$^3$, and 1057 mm$^3$. Due to the short interval of four days in the 10Gyd0d4 group, the impact of anti-PD-L1 on inhibiting tumor growth was exceedingly modest, as evidenced by a measured volume of 1125 mm$^3$ and a predicted volume of 1109 mm$^3$ on day 22. In contrast, the model prediction for the 15Gyd0d10 group revealed a large difference in tumor volumes with anti-PD-L1 (210 mm$^3$) compared to those without anti-PD-L1 (1104 mm$^3$) on day 22, resulting in a difference of 894 mm$^3$. A comparable disparity of 599 mm$^3$ was observed in the experimental data. These modeling results, which aligned with experimental observations, confirmed that our model effectively captured the disruption of key immunological processes occurring between days 1 and 10 when an additional radiation dose was applied. This was accomplished by accurately characterizing the distinct radiation responses of the two T cell subpopulations through parameters $\phi_1$ and $\phi_2$, as well as the conversion rate $\lambda$ between them. Specifically, $\phi_1$ was significantly lower than $\phi_2$, indicating that intratumoral T cells are more radio-resistant compared to the newly infiltrating T cells. According to our model and the fitted parameters, radiation recruited newly infiltrating T cells at an estimated rate of $\rho$ = 1.7. However, newly infiltrating T cells required time to convert into the more radio-resistant type. If additional radiation was delivered before this conversion was complete, then the more radio-sensitive new T cells were preferentially killed, which could adversely impact tumor control. In the experiment data, the 15Gyd0d10-CD8 group with anti-PD-L1 (1282 mm$^3$) differed from that without anti-PD-L1 (969 mm$^3$) in tumor volume on day 22 by 313 mm$^3$, which was estimated to be a difference of 17 mm$^3$ (913 mm$^3$ versus 930 mm$^3$). The modeling result appeared to differ from the actual value in this regime with the presence of anti-PD-L1 potentially due to the challenge of accurately predicting increased tumor volume control penalty when the anti-PD-L1 was present. However, our model showcases that

the depletion of T cells, due to anti-CD8 administration also accelerates tumor growth, even when fractions are properly spaced in the presence of a checkpoint inhibitor.

20Gyd0d10 and 40Gyd0: On day 22, the simulated differences in tumor volume were 134 mm$^3$ (92 mm$^3$ for 20Gyd0d10 with anti-PD-L1 versus 226 mm$^3$ for 20Gyd0d10 without anti-PD-L1) and 113 mm$^3$ (75 mm$^3$ for 40Gyd0 with anti-PD-L1 versus 188 mm$^3$ for 40Gyd0 without anti-PD-L1). These values approximated the observed differences of 298 mm$^3$ and 126 mm$^3$ in the experimental data. Nonetheless, we observed notable disparities between experimental outcomes and simulation results for these two treatment scenarios. Specifically, between days 2 and 16, there was a pronounced divergence between the blue and red lines. One likely explanation was that the small tumor volumes in these groups were represented with less precision than larger volumes observed in other groups. Another potential explanation was that one of our model equations was based on a modified LQ formulation, which was typically validated for lower, fractionated doses. At very high doses, such as 40 Gy, the predictive accuracy of the LQ model may have diminished. Moreover, since the PULSAR effect was evaluated as a relative metric, our primary focus was on overall tumor control achieved post-treatment than on transient fluctuations during the treatment course.

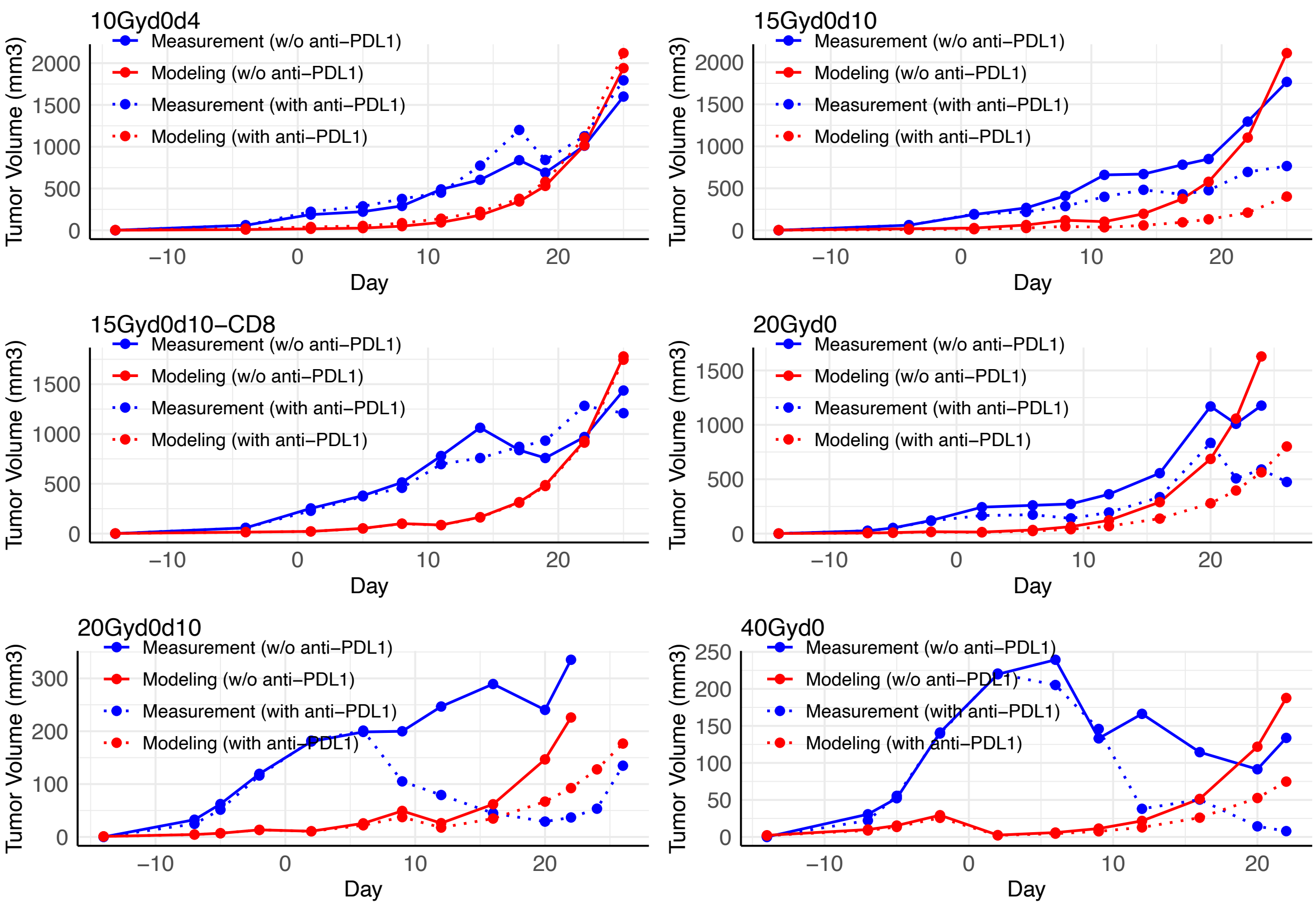

**Figure 3.** Experimental data (blue) in the testing data set of groups of different time spacing between the two fractions vs model-based tumor growth (red). In the above plots, the solid line represents radiotherapy only while the dashed line represents radiotherapy and immunotherapy combined. The 'Day' on the x-axis represents the day after the first RT if at least one radiation pulse is administered.

### *4.4. In-silico outcome prediction*

As described in Section 3.3, we simulated tumor growth for different treatment groups using the fitted parameters described in Section 4.2 and Table 2. Figure 4 shows the simulated tumor growth of twelve possible treatments, including a four-day interval (red), eight-day interval (blue), and twelve-day interval (green) of two radiation pulses, each of 15 Gy (top row) or 20 Gy (bottom row). For the 15 Gy scenario, the PULSAR effect was most significant when the two pulses were separated by 12 days compared to the counterparts of 4 days and 8 days. For the 20 Gy scenario, the overall tumor volume was lower due to the high dose. The PULSAR effect was also most noticeable when the two pulses were separated by 12 days. The results in Figure 4 exemplify the complex relationship between the optimal synergy and dose sequence (or fractionation) of radiotherapy.

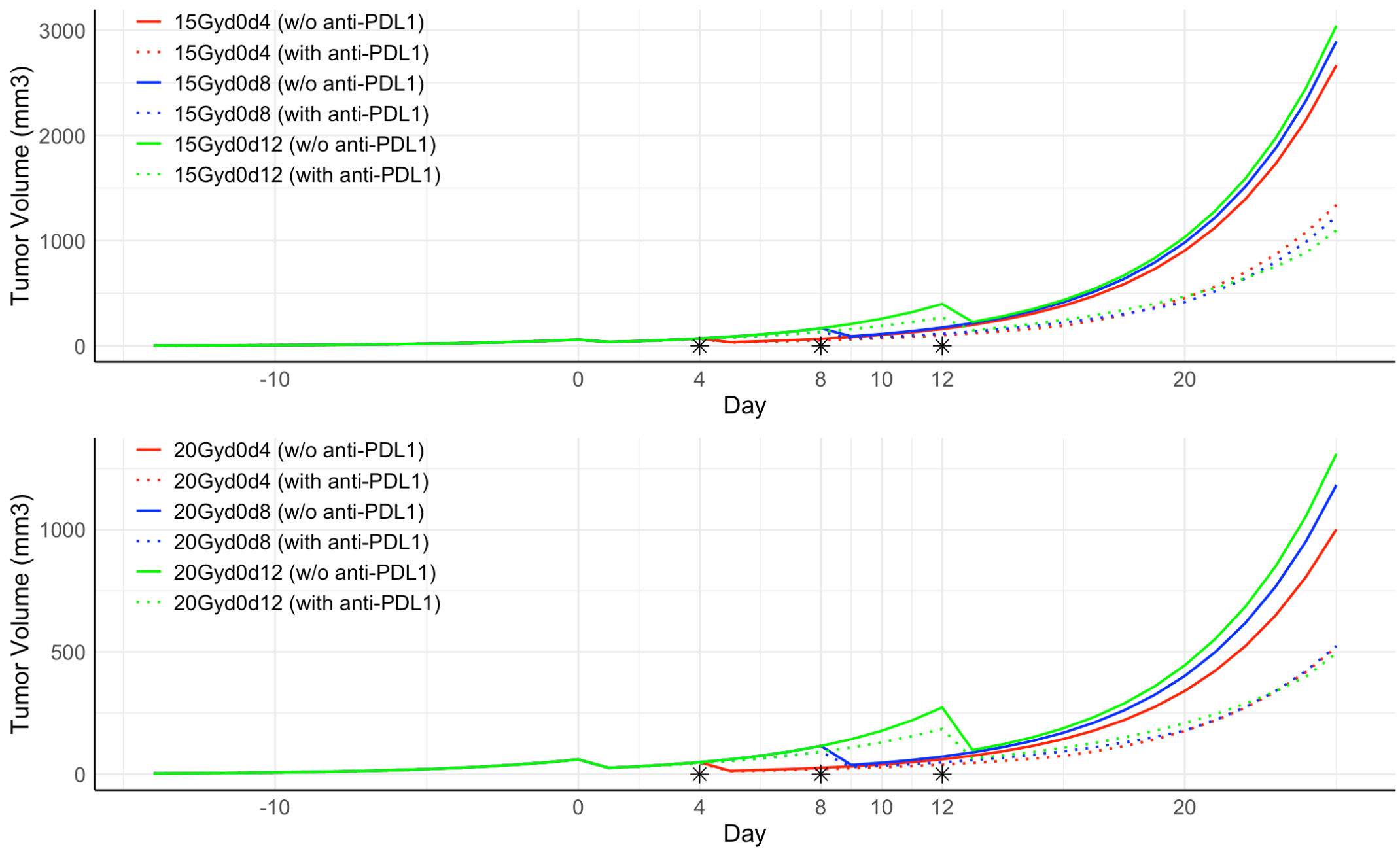

**Figure 4.** The simulated tumor growth of six possible treatments, including a four-day interval (red), eight-day interval (blue), and twelve-day interval (green) of two fractions each of 15 Gy (top) or each of 20 Gy (bottom). The '*' symbol denotes the day on which the second radiation treatment was administered in each treatment group.

## 5. Discussion

In this study, we constructed a mathematical framework for modeling the PULSAR effect utilizing a set of difference equations. This framework has potential value in illuminating the temporal dynamics of tumor control, shedding light on the combined effects of radiation and adaptive immune response, albeit in a qualitative or, at most, semi-quantitative manner. This is demonstrated in Figures 2 and 3 for various groups with distinct time intervals. Additionally, the model can support in silico studies and generate essential data to address additional research questions related to combined therapy.

While mathematical simplicity is intentionally attained with minimal components and parameters, the equations may appear unsophisticated and overly simplified. However, such fundamental models are frequently utilized in practical biological modeling, including tumor growth models, pharmacodynamics models, and the well-known LQ model in radiobiology. We also believe the chosen top-down approach is easier for future translation into clinical practice. Some parameters, such as $\tau, \phi_1, \phi_2, \lambda$, can be validated and fine-tuned through experiments, including flow cytometry and molecular imaging technologies. For example, optical imaging tools can be employed to investigate the dynamics of T cells, comprising their migration from lymph nodes to the targeted tumor via the circulatory system and subsequent infiltration.

The "increased radio-sensitivity", as defined in Eqs (2) and (3), is an essential component for fitting accuracy. The original linear-quadratic term is multiplied by the additional term for increased radio-sensitivity post-irradiation, involving parameters $\gamma$ and $\tau$ in a recursive manner. The cumulative effect decreases over time, lasting approximately ten days in correspondence with the decaying coefficient $\tau$. As shown in Figure 2, tumor control is more noticeable in group 10Gyd0d1 without anti-PDL1 (two 10 Gy pulses, day 0 and day 1), compared to group10Gyd0d10 without anti-PDL1 (two 10 Gy pulses, day 0 and day 10). However, two additional questions remain answered. First, is there an underlying radiobiological cause possibly linked to the increased radiosensitivity? Second, is the exponential decay of the cumulative effect of a single pulse associated with cell repair?

Another novel aspect of our study is the integration of T cell dynamics into radiation, including the differentiation and conversion of preexisting intratumoral and newly infiltrating T cells. Multiple investigations have highlighted the critical role played by the density and characteristics of tumor-infiltrating T cells in determining the clinical effectiveness of anti-PD-1/PD-L1 therapy. Arina et al. [5] reported that resident T cells present during radiation are more resistant to death and are important for overall tumor control. However, T cells in other compartments of the body exhibit lower resistance to radiation. While the authors did not administer a second radiation 24 hours post-initial dose, we find this aspect to be intriguing for exploring the synergistic potential between radiotherapy and immunotherapy. We speculate that newly infiltrated T cells may be more sensitive to the subsequent radiation pulse, given their phenotypic resemblance to T cells in other sites, which leads to the formulation in Eq (5). Without considering the immunomodulatory effect, the parameter $\tau$ results in reduced tumor control for a larger spacing between two pulses. However, such a pattern is altered after considering the temporal dynamic of T cells. When the two pulses are spaced ten days apart, the PULSAR effect is observed to be most significant compared to other spacing schemes (1 day, 4 days, and 20 days). Furthermore, the estimated parameters $\phi_1, \phi_2, \lambda$ support our speculation that T cell reprogramming is closely linked to the PULSAR effect. Different values of $\phi_1$ and $\phi_2$ reveal two types of T cells of different radiosensitivity. A subpopulation of T cells gradually acquires resistance and transforms into intratumoral T cells at a rate of $\lambda$. While we do not anticipate the proposed model to provide quantitative results, the investigation of these parameters increases our confidence in the validity of this approach.

Two plausible biological processes that may explain the maximum PULSAR effect around 10 days are briefly described below, with one process related to Treg. To prevent autoimmune reactions following radiation, Treg cells are recruited to the tumor site to suppress immune response through several signaling pathways, such as elevated expression of CTLA4 for suppressing inflammation response of antigen-presenting cells, cytokines IL10 for T helper cells, and TGF-β for cytotoxic T cells. Moreover, some Treg cells infiltrate the tumor bed and release type-I interferon gamma, which leads to the upregulation of PD-L1 on the surface of tumor cells. This process unfolds over several days following radiation. Another process may be related to the homing of T cells. In “hot” tumors, there is a robust immune response characterized by significant immune cell infiltration and activation of immune pathways. As a result, a single radiation dose may maximize synergy with immunotherapy, as initial priming of the immune response has already occurred due to preexisting tumor immunity. Conversely, “cold” tumors such as LLC in our study exhibit limited immune response, with minimal immune cell infiltration and reduced immune activity. Therefore, an initial priming dose of radiation is necessary, and the PULSAR effect manifests after approximately 10 days.

We present several limitations that will be addressed in future research. First, the model development was based on limited data, with only five to ten subjects per treatment group. Researchers should incorporate larger cohorts to enable more precise estimation of both within-group and between-group variability, thereby enhancing the model's accuracy. Also, while our model generally demonstrated robust performance, we observed substantial discrepancies in the 40Gyd0 and 20Gyd0d10 groups. These differences suggest that the model may be less accurate when applied to cases with small tumor volumes—likely due to increased uncertainty in parameter estimation for smaller tumors and higher radiation doses. Moreover, since our approach emphasizes the relative nature of the PULSAR effect by focusing on ultimate tumor control than transient fluctuations during treatment, the observed early-phase differences may be inherent to our evaluation strategy. Future work will entail to refine the model to better capture the dynamics of small tumor volumes and validate its performance across a broader range of treatment conditions. Additionally, we focused solely on the tumor size, lacking information on T cells related to migration and infiltration. A potential remedy involves conducting flow cytometry and optical imaging experiments to better understand the temporal behavior of T cells. Third, we focused on the PULSAR effect, not yet considering the risk of potential toxicities. As the combined radioimmunotherapy has the potential to result in delayed toxicities, the standard assessment models may not be appropriately suited.

## 6. Conclusions

The synergistic potential of combining radiotherapy and immunotherapy plays a unique role in cancer treatment. Here, we introduce a mathematical model based on multiple difference equations to investigate the PULSAR effect, which shows consistency with the experimental findings. An innovative aspect of our model is the incorporation of T cell dynamics alongside pulsed radiation, including the differentiation and conversion of both preexisting intratumoral and newly infiltrating T cells. This model provides a solid foundation for exploring the therapeutic synergy between radiotherapy and immune checkpoint blockade. It not only aids in determining the optimal dose and fractionation of radiotherapy for eliciting the optimal immune response but also offers insights into optimizing the dosing and sequencing of immune checkpoint inhibitors.

# Appendix

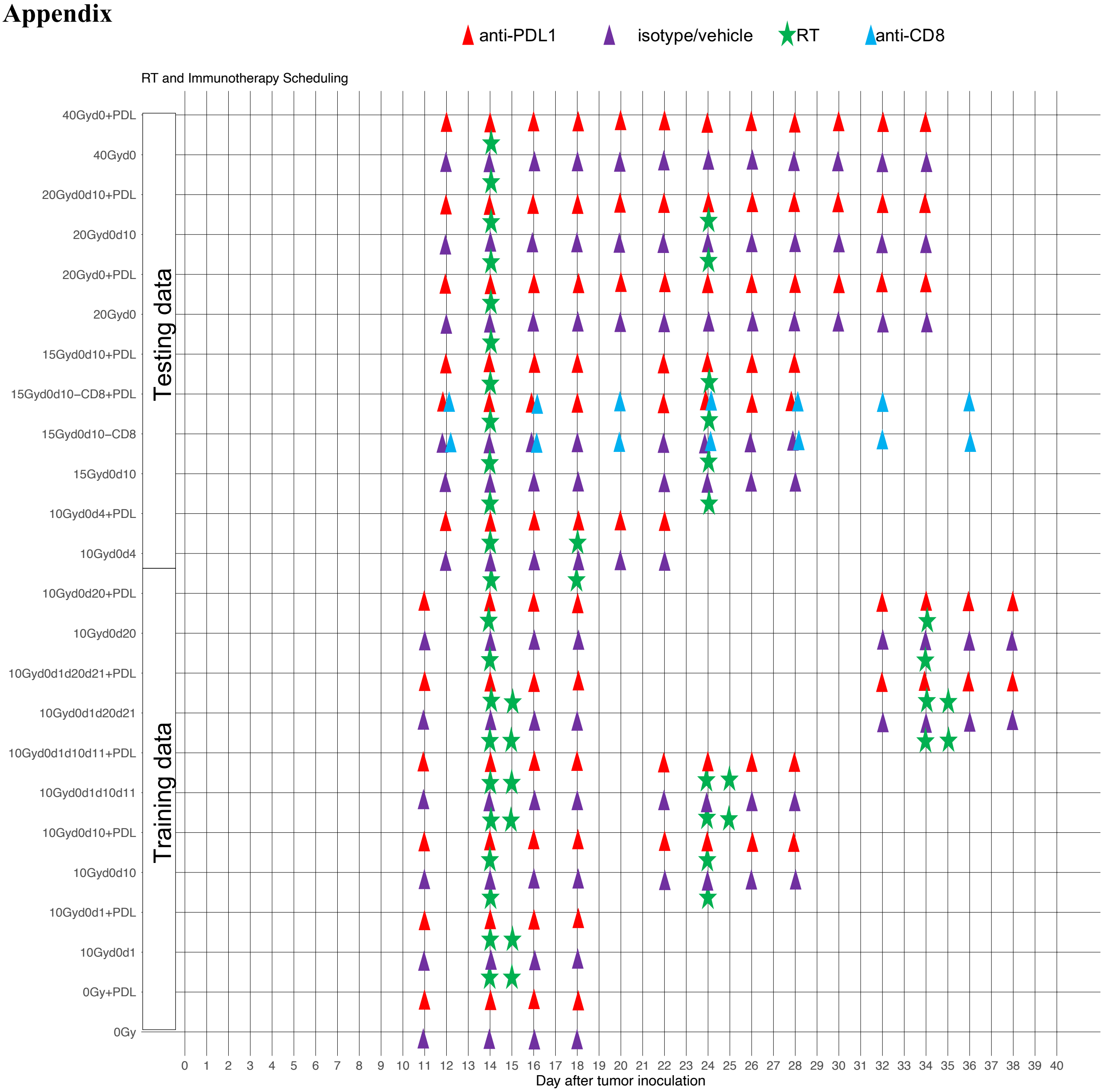

**Figure A1.** Radiotherapy and immunotherapy scheduling.